\documentclass[aip,amsmath,author-year,amssymb,reprint]{revtex4-1}
\pdfoutput=1

\usepackage{graphicx}
\usepackage{dcolumn}
\usepackage{bm}

\usepackage{multimedia}
\usepackage{chemfig}

\usepackage[utf8]{inputenc}
\usepackage[T1]{fontenc}
\usepackage{mathptmx}
\usepackage{etoolbox}

\usepackage{rotating}

\makeatletter
\def\@email#1#2{%
 \endgroup
 \patchcmd{\titleblock@produce}
  {\frontmatter@RRAPformat}
  {\frontmatter@RRAPformat{\produce@RRAP{*#1\href{mailto:#2}{#2}}}\frontmatter@RRAPformat}
  {}{}
}%
\makeatother
\begin{document}


\title[Stochastic method for isotope labeling systems]{Stochastic simulation algorithm for isotope-based dynamic flux analysis}


\author{Quentin Thommen}
\affiliation{Univ. Lille, CNRS, Inserm, CHU Lille, Institut Pasteur de Lille, UMR9020-U1277 -
CANTHER - Cancer Heterogeneity Plasticity and Resistance to Therapies, F-59000}
\email{quentin.thommen@univ-lille.fr}
\author{Julien Hurbain}

\author{Benjamin Pfeuty}
\affiliation{Univ. Lille, CNRS, UMR 8523 - PhLAM - Physique des Lasers Atomes et Mol\'ecules, F-59000 Lille, France}

\date{\today}

\begin{abstract}
  Carbon isotope labeling method is a standard metabolic engineering tool for flux quantification in living cells.
  To cope with the high dimensionality of isotope labeling systems, diverse algorithms have been developed to reduce the number of variables or operations in metabolic flux analysis (MFA), but lacks generalizability to non-stationary metabolic conditions. 
  In this study, we present a stochastic simulation algorithm (SSA) derived from the chemical master equation of the isotope labeling system.
  This algorithm allows to compute the time evolution of isotopomer concentrations in non-stationary conditions, with the valuable property that computational time does not scale with the number of isotopomers.
  The efficiency and limitations of the algorithm is benchmarked for the forward and inverse problems of 13C-DMFA in the pentose phosphate pathways. 
  Overall, SSA constitute an alternative class to deterministic approaches for metabolic flux analysis that is well adapted to comprehensive dataset including parallel labeling experiments, and whose limitations associated to the sampling size can be overcome by using Monte Carlo sampling approaches.\\

\vspace{1cm}


\end{abstract}
      
\maketitle

\section{Introduction}


Isotope tracing experiments have been developed to quantify fluxes in biochemical networks~\citep{stephanopoulos1999metabolic}.
A typical carbon-13 labeling experiment metabolizes a labeled substrate, such as [1- 13C]glucose, tracks the propagation of the label on metabolites by nuclear magnetic resonance (NMR) or mass spectrometry (MS) methods and estimates metabolic fluxes by various methods including 13C-MFA~\citep{niedenfuhr2015,allen2020,antoniewicz2021guide}.
Despite its limitations, 13C-MFA remains the gold standard method in metabolic engineering for accurate and precise quantification of fluxes in living cells \citep{crown2013parallel}.
 Currently, the most efficient algorithms are all based on an advanced decomposition method using elementary metabolic units (EMUs) developed in 2007 by Antoniewicz et al \citep{antoniewicz2007emu}.
 Nevertheless, one of the limitations of the classical metabolic flux analysis (MFA) method is the requirement of a metabolic isotopic steady state.
Flux analysis methods that focus on estimating non-stationary metabolic fluxes are referred to as dynamic MFA (DMFA)~\citep{leighty2011DMFA}, or 13C dynamic MFA methods (13C-DMFA) methods~\citep{antoniewicz2015}.
Despite pioneering works~\citep{antoniewicz2007DMFA,Wahl2008} initiated more than one decade ago, little progress has been made since \citep{antoniewicz2021guide}.
Current computational methods use a deterministic modeling framework by solving EMU balance rate equations where dynamic flux parameters are modeled with B-splines~\citep{quek2020,ohno2020}.
Computational tractability of such method depends on the EMU system size that can be very large due to the interplay of elaborated labeling protocols \citep{lewis2014tracing,antoniewicz2015parallel,Jacobson2019,Dong2019,allen2020} and complex bibi reactions~\citep{selivanov2004}.

In this paper, we present a different class of method that simulates isotope propagation in non-stationary metabolic systems by a Stochastic Simulation Algorithm (SSA).
We test the method in the metabolic subsystem comprising glycolytic and PPP pathways where complex carbon rearrangements occur due to bibi reactions in the nonoxidative PPP and where 13C labeling have been extensively applied to infer metabolic flux~\citep{kuehne2015,bouzier2015,creek2015,diaz2016,lee2019}.
The main idea is to represent the population of isotopomers of a chemical species by a sample of finite size, proportional to the species concentration, and to use the standard rules of stochastic chemical kinetics to propagate the marker.
When a reaction occurs, the isotopomers associated to the reactants are randomly selected in the corresponding samples, the rearrangement is performed, and the products are added to the corresponding samples.
The algorithm somehow mimics the discrete and stochastic processes of enzymatic reactions as it occurs in cells, but remains restricted to a small sample of metabolite species for the sake of computational efficiency.
At each time step, the samples represent the population of the isotopomers of each variable from which one can compute mass isotopomer distribution for comparing with experimental data.
The proposed algorithm is simple to implement, fast, visual, and above all its computation time depends very little on the chain length, which makes it an algorithm also adapted to parallel labeling (13C, 2H, 15N, 18O, etc).

\section{Results}

\subsection{Stochastic Simulation Algorithm (SSA)}

\begin{figure}
	\centering
	\includegraphics[width=7cm]{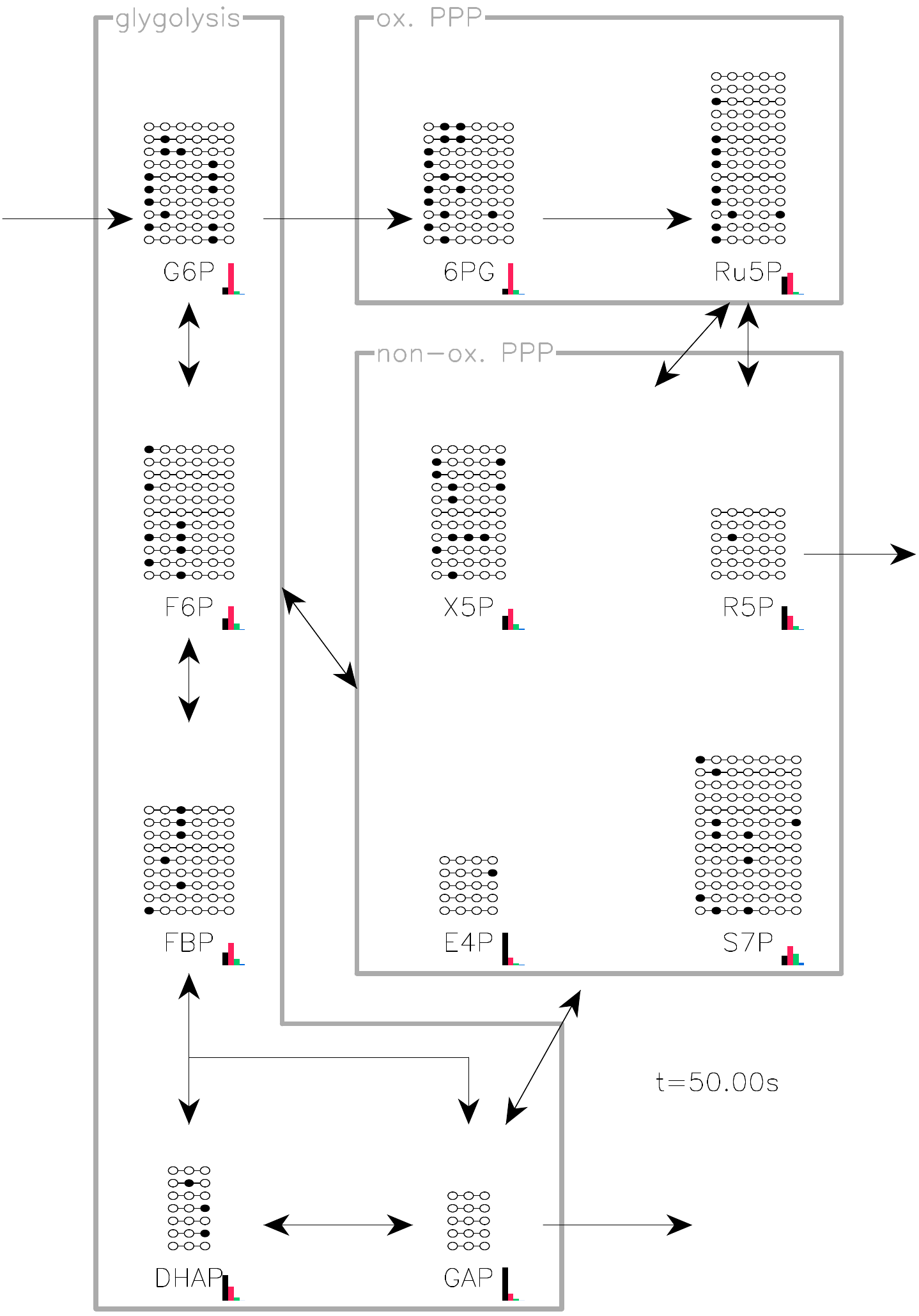}


        \caption{\label{fig:01} Atom rearrangements in metabolic reactions.
          The upper glycolytic pathway supplemented by the pentose phosphate pathway provides an example of isotope labeling network. For each chemical species (G6P, F6P, FBP, DHAP, GAP 6PG, Ru5P, X5P, R5P, E4P, S7P)  subsamples of the isotopomer samples are displayed as a chain of unlabeled (open circle) and  labeled carbons (close circle). The number of displayed carbons chains is proportional to the species concentration.  Mass isopotomer histogram is also displayed (m+0 in black, m+1 in red, m+2 in green, and m+3 in blue). The Figure illustrates the configuration 50~s after the labeling introduction that also corresponds to the perturbation of the metabolic system. The associated video provides a full dynamical picture.              }
              \label{fig:video}
\end{figure}

The propagation of labeled atoms through a biochemical network is here described by a sampling approach. The representation of the isotopomer distribution of each chemical species in the network is computed using a finite sample size proportional to its concentration. A user defined parameter $\Omega$ corresponds to a reference concentration. For example, a value of $\Omega=1000$ c/$\mu$M indicates that a concentration of 1~$\mu$M is represented by 1000 copies of the chemical species, each copy corresponding to a different isotopomer. 

The fluxes of chemical reactions are determined by mathematical functions that can be either linked to the  species concentrations in the framework of chemical kinetics, or described by phenomenological functions depending on time, or by constant functions in the case of stationary flux condition.
The flux value determines the time interval between two occurrences of the corresponding chemical reactions. When one occurs, the reactants are taken randomly from the corresponding samples, the rearrangement of the atoms is done according to the reaction's rule, and the products are added to the corresponding samples. In this way, the labeling propagates through the chemical reaction network; at a given date, the sample of each species is populated with different isotopomers and represents the isotopomer distribution.

Such rules are formalized within the framework of the chemical master equation once two new tools are defined, the isotopomer index and addressing operators (sec.~3.1). Chemical master equation  describes the temporal evolution of the isotopomer fraction. From the chemical master equation, one can derive both a deterministic simulation algorithm (DSA) (see Sec.~3.2) and a stochastic simulation algorithm (SSA) (see Sec.~3.3). The DSA is not an efficient algorithm since it has as many variables as possible isotopomer, it is a "brute force" algorithm serving here as a control for the SSA outputs.

An example of stochastic simulation is given in Figure~\ref{fig:video} and for the upper glycolytic pathway combined with the pentose phosphate pathway. 
To determine the fluxes, the mass action law is here used with unitary kinetic parameters (Table~1).
At the initial time, the metabolic system is fed with labeled glucose (50\% of [1-13 C]glucose and 50\% of [2-13 C]glucose in~\citep{kuehne2015}), and at the same time, is perturbed by a two-fold increase of the glucose intake rate.
If $\Omega=100$ c/$\mu$M is used for SSA, the Figure~\ref{fig:video} (and the corresponding video) only represents one element out of 20 from each sample, for the sake of visualization.

\begin{figure}
	\centering
	\includegraphics[width=8cm]{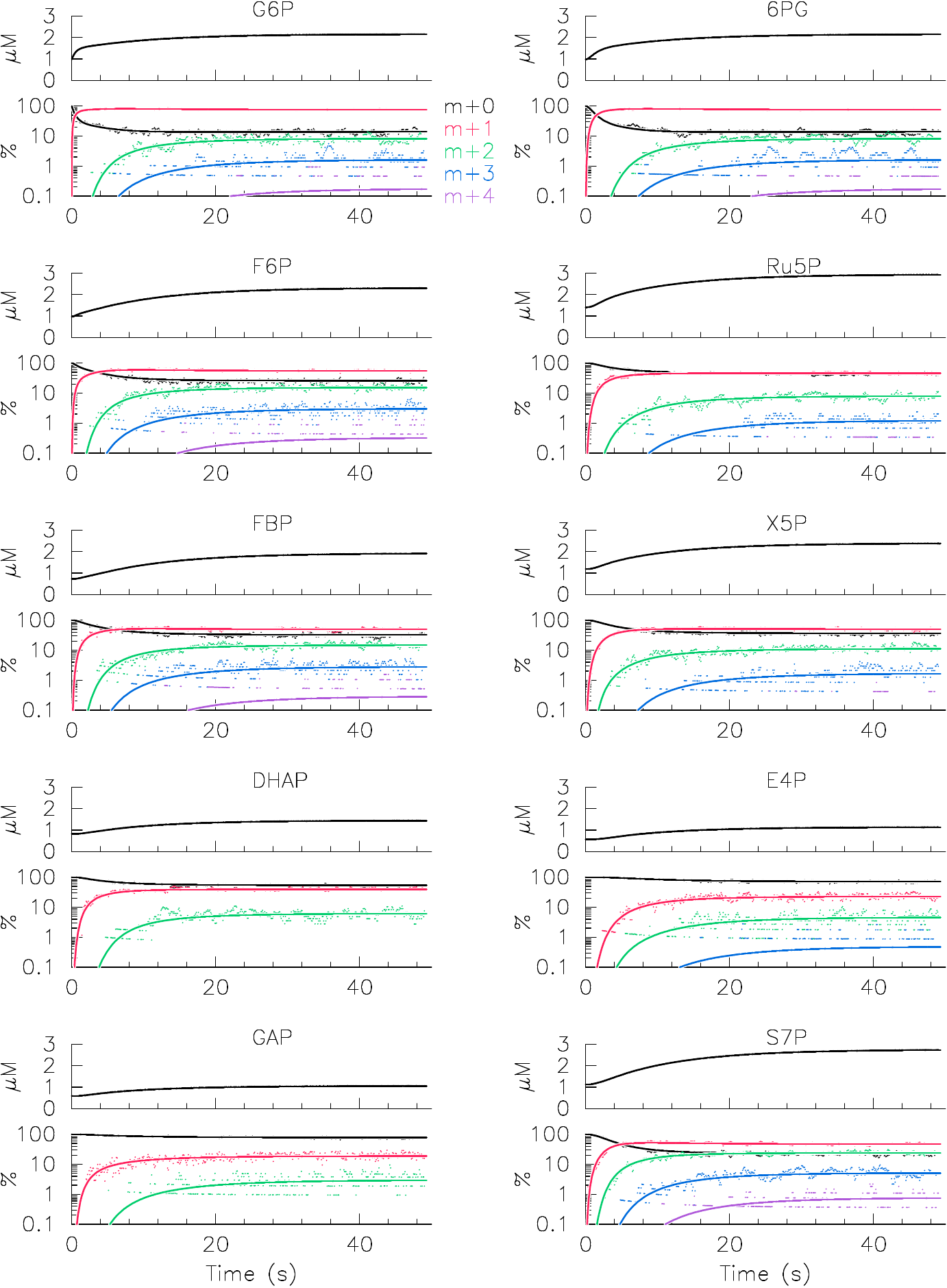}
        \caption{Concentration and mass isotopomer dynamics in nonstationary conditions.
          Stochastic trajectories computed with SSA (dots) are compared with deterministic trajectories computed with DSA for control (solid lines), and corresponds to calculation presented in Figure~1 (same network, same condition). For each chemical species, the upper plot displays the concentration in $\mu$M whereas the bottom plot displays the mass isotopomers in percent (m+0 in black, m+1 in red, m+2 in green, and m+3 in blue, m+4 violet). 
}
              \label{fig:trace}

      \end{figure}

The Figure~\ref{fig:trace} represents the evolution of the concentration and mass isotopomer obtained with the SSA (point) and the DSA (continuous line), thus depicting the accurate trends of isotopomer trajectories generated with SSA. The stochastic fluctuations of the mass isotopomers induced by the SSA are only due to the random selection of the reagents in the sample. The variance of these fluctuations is thus equal to the $\Omega$ profile and the mass isotopomer concentration. The determination is thus all the more precise as the mass isotopomers are abundant.
It is thus possible to reduce these fluctuations in two different ways, either by increasing the value of $\Omega$, or by proceeding to a temporal smoothing of the stochastic evolution.

\subsection{SSA Computational Performance}

SSA computation time depends on both the number of chemical reactions and the execution time of each reaction. As an example, the computation cost necessary to simulate data of the Figure~\ref{fig:video} corresponds to 224 463 reactions carried out in 62~ms for the SSA, and 2892 right-hand-side evaluation in 2100~ms for the DSA (advanced Runge-Kutta-Fehlberg method is used) using a Intel(R) Core(TM) i5-6300U CPU at 2.40GHz without parallelization.
In the SSA, the number of chemical reaction occurrences can be approximated by the product $T\,v\,\Omega\,N$ where $N$ is the number of chemical reactions in the network, $v$ the typical flux values and $T$ the time interval.
The number of reactions does not depend on the number of isotopomer per species or, equivalently, on the chain length representing the chemical species.
The time to perform a reaction depends only slightly on the chain length $l$ thus almost not depend on the number of isotopomer. In our implementation, the computation time for one reaction varies as $1+l/15$; so when $l$ goes from $6$ to $18$ (e.g., C6 to C6H12), the computation time increase by less than 60\% whereas the number of isotopomer is multiplied by $2^{12}=4096$. 
This is why the SSA algorithm is well adapted to cross-labeling, \textit{e.g.} hydrogen carbon, leading to longer chain lengths and thus to a higher combinatoriality.

\subsection{SSA-13C DMFA}

\begin{figure}
	\centering
	\includegraphics[width=8cm]{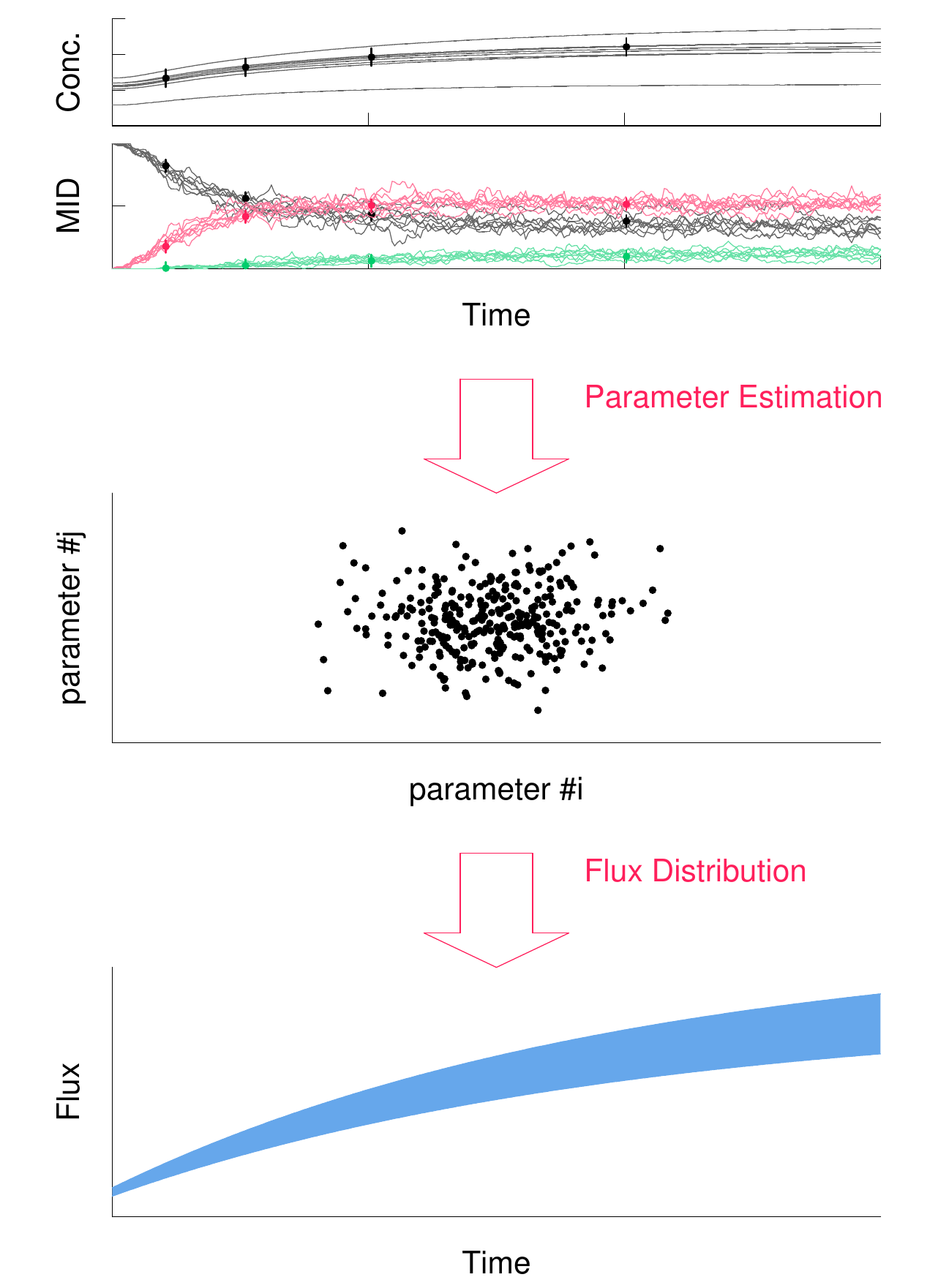}
        \caption{13C-DMFA generic workflow using SSA.
          Metabolite concentration as well as mass isotopomer distribution (MID) are the targets of parameter estimation procedure. Parameter sensitivity analysis provides  a list of points in the parameter space that accurately describes targets. The range of flux dynamics is then computed from the parameter value distribution.       
}
              \label{fig:FIT_METHODO}

      \end{figure}

To further test the SSA, we implemented it in a 13C-DMFA procedure.
A general scheme of the procedure is shown in Figure~\ref{fig:FIT_METHODO}.
A series of measurements concerning metabolite concentrations and mass isotopomer distributions (MID) with known associated experimental errors is the target of an optimization procedure. 
The aim is to fit these data with a kinetic model based on mass action laws used for Figure 1 (Table 1).
The flux dynamics therefore depend on the kinetic parameters of the reaction laws.
Instead of a kinetic model, we could also use the stochiometric model supplemented with parameterized time functions to describe the flux. The parameter space of the model is then explored to identify the sets of parameters consistent with the target experimental data, taking into account the existing uncertainties (see ~\citep{valderrama2019} for a review of standard method). Once the exploration is completed, the dynamics of metabolic fluxes are computed for each selected parameter set, which can be represented as a confidence region for flux trajectories. 

\begin{figure}
	\centering
	\includegraphics[width=8cm]{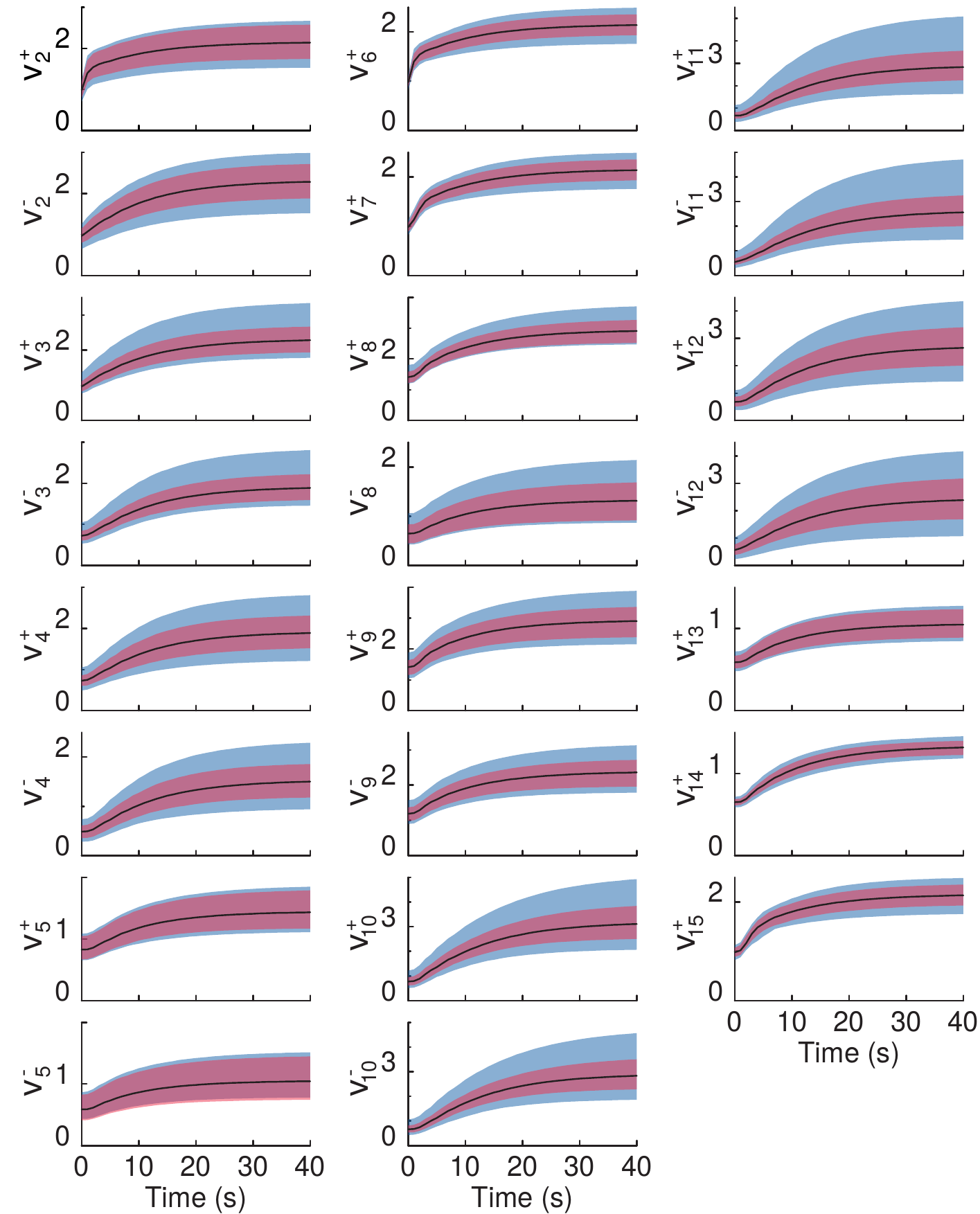}
        \caption{Directional flux areas compatible with mass isotopomer targeting.
          The areas are filled with a color code.
          Blue  areas correspond fit strategy 1; red, to fit strategy 2. Solid black lines correspond to the exact solution.}
              \label{fig:FIT}

      \end{figure}

Here, target datasets were generated for concentrations and mass isotopomers from the DSA at ${2,5,10,20,30,40}$s in the same condition as in Figure~\ref{fig:trace}.
Then the SSA, with $\Omega=200$ here, is used to compute the fitness score from target dataset and kinetic parameter set.
Two classes of experimental measurements are considered.
In a first strategy, only the mass isotopomers $m+0 \dots m+3$ are targeted with an error of 5\% (no data provided for the concentrations).
In a second strategy, concentration data are also included and targeted with an error of 0.25 $\mu$M. These errors mimic typical experimental and measurement uncertainties.
 A parameter set is here kept and said to be consistent with target dataset for a chi-square per degree of freedom (i.e., fitness score) remains below unity. 
Here, a parameter sensitivity analysis computes the parameter ranges, one by one, to illustrate the procedure; the input flux is assumed to be known.
For each strategy, the selected parameter sets are finally used to compute the dispersion of the reaction fluxes.
As expected, the areas of flux trajectories comprise the exact solution and are reduced when adding concentration data (Figure~\ref{fig:FIT}).

\section{Theory}

\subsection{Chemical master equation model for isotopic labeling networks}

In a network of (bio)chemical reactions, the temporal evolution of the state probabilities is described by the chemical master equation (CME) through a general formalism~\citep{Gillespie1992}.
Deterministic kinetic rate equations, on the one hand, can be derived from the first moments of the probability distribution and allow for a thorough analysis of the network dynamics by various analytical techniques ~\citep{Thompson2002}.
The probabilistic features of the dynamics such as bimodal distributions or coefficients of variation, on the other hand, can be investigated with stochastic implementation of the CME through well-established stochastic simulation algorithms \citep{gillespie1977exact,gillespie2001approximate}.

\subsubsection{Isotopomer index and addressing operators}

A chemical reaction network such as the one depicted in Figure~\ref{fig:video} is defined by $K$ reactions between $M$ chemical species whose concentrations are denoted by $S_{m}$ with $m\in[1,M]$.
The labeling states of the species $S_{m}$ is an ordered sequence $(s_{m,1},s_{m,2},\dots,s_{m,l_m})$ of length denoted $l_m$ made of elements $s_{m,i}\in[0,q-1]$.  
The species $S_{m}$ has therefore $L_m=q^{l_m}$ different labeling states or positional isotopomer indexed by $n_m=\sum_{i=1}^{l_m}\, s_{m,i}\,q^{i-1}$, also noted $n_m=(s_{m,1},s_{m,2},\dots,s_{m,l_m})_q$ and called the isotopomer index.
A similar approach restricted to $q=2$ has already been introduced to describe isotopomer distribution vectors~\citep{schmidt1997idv}.   
In the case 13C-labeling, each carbon may be in two different states (\textit{i.e.}, $q=2$) and the sequence $(0,0,1,0,0,0,0)$ for S7P indicates that 13C label is in third carbon position and corresponds to the labeling state number 4, the S7P species has therefore $2^7$ different labeling states.

If the permutation rule is known, one can define addressing operators that compute the isotopomer index of the products from the isotopomer index of the reactants, and \textit{vice versa} for each reaction of the network. The addressing operator forms an alternative to atom mapping matrices defined by Zupke \textit{et al.}~\citep{zupke1994}.
In the case of the reaction mediated by transaldolase (reaction number 11 in Table~1), the addressing operators
\begin{align}
\sigma_{\mathrm{F6P}}(n_{\mathrm{S7P}},n_{\mathrm{GAP}})=&(s_{a,1},s_{a,2},s_{a,3},s_{b,1},s_{b,2},s_{b,3})_q \\
\sigma_{\mathrm{E4P}}(n_{\mathrm{S7P}},n_{\mathrm{GAP}})=&(s_{a,4},s_{a,5},s_{a,6},s_{a,7})_q,
\end{align}
compute the product index from reactant index
$n_{\mathrm{S7P}}=(s_{a,1},\dots,s_{a,7})_q$ and
$n_{\mathrm{GAP}}=(s_{b,1},s_{b,2},s_{b,3})_q$.
In the same manner, 
\begin{align}
\sigma_{\mathrm{S7P}}(n_{\mathrm{F6P}},n_{\mathrm{E4P}})=&(s_{c,1},s_{c,2},s_{c,3},s_{d,1},s_{d,2},s_{d,3},s_{d,4})_q\\
\sigma_{\mathrm{GAP}}(n_{\mathrm{F6P}},n_{\mathrm{E4P}})=&(s_{c,4},s_{c,5},s_{c,6})_q
\end{align}
compute the reactant index from product index
$n_{\mathrm{F6P}}=(s_{c,1},\dots,s_{c,6})_q$ and
$n_{\mathrm{E4P}}=(s_{d,1},\dots,s_{d,4})_q$. 
Therefore, in the context of a 13C labeling ($q=2$), the reaction between a doubly labeled S7P (1000100) and a simply labeled GAP (001)  -- \textit{i.e.} $n_{\mathrm{S7P}}=17$ and $n_{\mathrm{GAP}}=4$ --  produces an F6P (100001) and an E4P (0100) -- \textit{i.e.} $n_{\mathrm{F6P}}=33$ and $n_{\mathrm{E4P}}=2$.

\subsubsection{Chemical master equation description}

The chemical master equation (CME) is a general and accurate formalism to describe the stochastic dynamics in (bio)chemical reaction networks \citep{gillespie2000}.
This formalism can be easily extended to also describe the stochastic dynamics of labeling states of chemical species. 
With above notations for isotopomer index and addressing operators, the probabilistic dynamics in isotope labeling network can naturally be described in the CME framework.
The chemical species with the largest number of isotopomers determines the size $N$ of the state space ($N=2^7$ in the example used in Figure~\ref{fig:video}).
The state of the whole system is therefore described by a $M \times N$ integer matrix $\mathbf{\omega}$: $\omega_{m,n}$ indicating the number of $m$th species in the $n$th labeling state. The total number of the $m$th species is noted $\Omega\,S_m=\sum_n\,\omega_{m,n}$ where $\Omega$ is a volume (involved as a scaling factor) and $S_m$ is a concentration.  The probability that the internal sequence of the $m$th species corresponds to the $n$th isotopomer  is denoted $\rho_{m,n}=\omega_{m,n}\,/\,\Omega\,S_m$. The formalism of the CME describes the temporal evolution of the probability of the system to be in the state $\mathbf{\omega}$, noted $\mathcal{P}_{\mathbf{\omega}}(t)$.

The chemical reactions that define the network are characterized by both a concentration-dependent flux of reagents $v_{k}$ with $k\in[0,K]$ and a permutation rule between the position of labeled atoms of reactants and products. 
Reactions are distinguished depending on their input, output, or internal position in the network. 
For instance, the network depicted in Figure~\ref{fig:video} has one input reaction, 3 output reactions and 11 internal reactions. 
As seen latter on, input reactions always require a particular consideration since the reactant is not modified.   
For keeping notations simple, we restrict to Bi Bi reactions of the type $A+B\rightarrow C+D$ where $A,B,C,D$ are either chemical species or empty sets.  
In this case, the CME reads,
\begin{widetext}
	\begin{align}
	\label{eq:master_eq}
	\frac{d}{dt}\mathcal{P}_{\mathbf{\omega}}(t)=&\sum_{k=1}^{K_n} \Omega\;v_k\left( S \right)\,
	\sum_{n,n'} \left[ \mathbb{E}_{a_k,n}^+\,\mathbb{E}_{b_k,n'}^+\,\mathbb{E}_{c_k,\sigma_{c_k}(n,n')}^-\,\mathbb{E}_{d_k,\sigma_{d_k}(n,n')}^- -1
	\right] \rho_{a_k,n}\,\rho_{b_k,n'} \mathcal{P}_{\mathbf{\omega}}(t) \\
	&+\sum_{k=K_n+1}^{K_n+K_i} \Omega\;v_k\left( S \right)\,
	\sum_{n} I_{c_k,n}(t)\;\left[ \mathbb{E}_{c_k,n}^- -1 \right] \; \mathcal{P}_{\mathbf{\omega}}(t) \nonumber
	\end{align} 
\end{widetext}
The integers ${a_k,b_k,c_k,d_k}$ correspond to the indices of the species ${A,B,C,D}$ of the $k$th reaction of type $A+B\rightarrow C+D$, the integer being null in the case of an empty set. The writing uses scale operators $\mathbb{E}_{m,n}^\pm$~\citep{van1992} :
\begin{widetext}
	\begin{equation}
	\label{eq:step_op}
	\mathbb{E}_{m,n}^\pm \;\rho_{m_1,n_1}\,\rho_{m_2,n_2}\,\mathcal{P}_{\mathbf{\omega}}(t)=
	\left[\rho_{m_1,n_1}\pm\frac{\delta_{m,m_1}\delta_{n,n_1}}{\Omega\,S_{m_1}}\right]
	\left[\rho_{m_2,n_2}\pm\frac{\delta_{m,m_2}\delta_{n,n_2}}{\Omega\,S_{m_2}}\right]
	\,\mathcal{P}_{\mathbf{\omega}\pm\mathbf{E}_{m,n}}(t)
	\end{equation}
\end{widetext}
where $\mathbf{E}_{m,n}$ is a matrix of the canonical base (only the element at the intersection of row $m$ and column $n$ is non-zero and is unity), and $\delta_{i,j}$ the Kronecker symbol (unity if indexes are equal, zero either).
The $K_n$ first reactions concern internal and output reactions whereas the remaining $K_I$ concerns input reactions ($\emptyset \rightarrow C$).
In this later case, $I_{c_k,n}$ is the fixed probability to have an isotopomer $n$ of the input species $c_k$. 


\subsection{Derived Deterministic Simulation Algorithm (DSA)}

The CME ({Eq.~\ref{eq:master_eq}}) can be approximated in the large size limit $\Omega\rightarrow\infty$ by deterministic rate equation dynamics.
The probability of each internal sequence is denoted by $\rho_{m,n}$ such that $S_{m,n}=S_{m}\,\rho_{m,n}$ describes the concentration of $n$th-isotopomer of the $m$th species.  
The time evolution of isotopomer concentrations is governed by the distribution rules specific to each reaction and is formalized mathematically by a permutation of the concatenated internal sequence between the reagents and the products.
The deterministic system evolves according to the ordinary differential equations,
\begin{align}
\label{eq:determ0}
\frac{d}{dt}S_{m,n} &= \sum_{k=1}^{K}N_{m,k}\ v_{k}(S)\; \Phi^k_{m,n}(\rho) \,,
\end{align}
where $N$ denotes the stoichiometry matrix, $v_{k}\; k\in[0,K]$ the concentration-dependent flux of reagents, and $\Phi^k_{m,n}(\rho)$ the flux fraction describing the permutation rules of chemical reaction satisfying $\sum_n \Phi^k_{m,n}(\rho)=1$.
Algorithm to simulate {Eq.~\ref{eq:determ0}} with standard Runge-Kutta-Fehlberg method of order 5 with adaptive step is called Deterministic Simulation Algorithm (DSA).

Let us first consider the internal reactions restricted to Bi Bi reactions of the form $A+B \rightarrow C+D$. 
The reaction is characterized by the reordering of atom position defining addressing operations of the products according to the indices of the reagents.
The operator $\sigma_{A}(n_{c},n_{d})$ gives the isotopomer index of the $A$ species that produce $C$ and $D$ of isotopomer index $n_c$ and $n_d$, respectively.
In that case, the reaction index $k$ is omitted and the flux fraction reads,
\begin{align}
\Phi_{a,n_a}(\rho)= & \rho_{a,n_a} \\
\Phi_{b,n_b}(\rho)= & \rho_{b,n_b}  \\
\Phi_{c,n_c}(\rho)= & \sum_{n_{d}=0}^{L_{d}-1}\rho_{a,\sigma_{A}(n_{c},n_{d})}\;\rho_{b,\sigma_{B}(n_{c},n_{d})}  \\
\Phi_{d,n_d}(\rho)= & \sum_{n_{c}=0}^{L_{c}-1}\rho_{a,\sigma_{A}(n_{c},n_{d})}\;\rho_{b,\sigma_{B}(n_{c},n_{d})}  
\end{align}
where $L_c$ and $L_d$ are the number of isotopomers of $C$ and $D$ species.
If $B$ is an empty set then $\rho_{b,x}=1$, and if $D$ is an empty set {Eq.~(11)} is useless.
If the reaction is an output reaction, then $C$ and $D$ are empty sets, $\Phi_{a,n_a}(\rho)$ and $\Phi_{b,n_b}(\rho)$ are computed with the same rules as internal reactions. 
As mentioned, the input reactions must be treated separately since the reactants are not variables but parameters. 
We consider here input reactions of simple form $\emptyset \rightarrow C$ and we note $I_n$ the probability of synthesis of the species $C$ in the state $n$, thus $\Phi_{c,n_c}(\rho)=I_n$ in this case.

Alternatively, the dynamical system ({Eq.~\ref{eq:determ0}}) can also be written as
\begin{align}
\label{eq:determ}
\frac{d}{dt}S_m&=\sum_{k=1}^{K}N_{m,k}\ v_{k}(S)\quad m\in[0,M] \\
S_m\,\frac{d}{dt}\rho_{m,n} &= \sum_{k=1}^{K}N_{m,k}\ v_{k}(S)\;\left( \Phi^k_{m,n}(\rho) - \rho_{m,n}\right) 
\end{align}
The first equation describes the time evolution of species concentrations while the second equation describes the time evolution of the fraction of different isotopomers.
This additional equation highlights the key role of the concentrations $S_m$ in the timescale of changes in isotopomer distribution: higher concentration values lead to slower evolution of isotopomer distributions.

The permutation rules defined in $\Phi$ may be easily extended to more complex reactions. 
In the framework developed here, they only depend on the permutation relations and not on the mathematical forms of concentration-dependent flux, because we assume that the internal modification does not impact the reaction rate. 
If the construction rules are simple to establish and to implement in a numerical code, the computation time of the flux vector of the dynamic system (the right-hand side term of {Eq.~\ref{eq:determ0}}) increases significantly with the length of the sequences and the number of isotopomers, because of the many summations of terms.
Moreover, this implementation computes the evolution of all possible isotopomers while the experimental labeling used nowadays generates only a small subset of the possible isotopomers~\citep{metallo2009}. 
The deterministic system therefore requires a large number of unnecessary calculations even with an optimized implementation.
It nevertheless serves as a useful benchmark to check the relative accuracy and efficiency of other methods.

\subsection{Derived Stochastic Simulation Algorithm (SSA)}

The CME is in fact a continuous-time approximation of discrete time stochastic processes. 
Stochastic algorithms are often used to simulate the molecular dynamics in chemical reaction networks and capture the statistical and temporal features of fluctuations.
In the case of the above CME ({Eq.~\ref{eq:master_eq}}), time evolution of isotopomer distribution can also be simulated by a stochastic Monte-Carlo algorithm based on the next reaction methods~\citep{gibson2000}, here called Stochastic Simulation Algorithm (SSA).
Each chemical species is represented by a finite sample of isotopomers (Figure~\ref{fig:video}) where the sampling size is proportional to the concentration of the corresponding chemical species.

The sample size of the variable $m$ is $\Omega\,S_m$ where $\Omega$ is a volume.
The occurrence of a chemical reaction is determined by the standard next reaction methods that we have adapted, SSA is summarized as: 
\begin{itemize}
	\item[Init:]  Compute the reaction time for all reactions
	\begin{equation}
	\label{eq:tau}
	\tau_k=\frac{1}{\Omega\;v_k\left( S \right) }
	\end{equation}
	\item[Step 1:] Find the smallest reaction time $\tau_{k'}=\mathrm{min}(\tau_k)$ and do reaction $k'$ by randomly picking the reagents from their samples  and synthesizing the products following the permutation rule of the reaction;
	\item[Step 2:] Increment time $t$ by $\tau_{k'}$ and compute a next time for reaction $k$;
	\item[Step 3:] Adjust the set of reaction times to account for sample size variation induced by reaction $k'$
	\[ \tau_k \leftarrow \frac{v_{k,\mathrm{old}}}{v_{k,\mathrm{new}}}\left(\tau_k-t \right) + t \]
	and iterate to Step~1
\end{itemize}
In this sequential process, each stochastic occurrence of a chemical reaction induces discrete changes in the number of species and of isotopomers associated to each chemical species, which results in stochastic fluctuations of both species concentrations and isotopomer distribution.

Contrary to the common use of stochastic simulation algorithms for chemical reaction networks, the $\Omega$ value does not have to represent the real number of molecules for a reference concentration because the algorithm considers mainly the propagation of marked atoms and not the stochastic fluctuations of the chemical reactions linked to the finite number of copies.
In the context of metabolic networks, fluctuation of the reaction times $\tau_k$  are indeed rarely relevant. Because of the high copy number of metabolites, numerous reactions occur and  fluctuations of the reaction times $\tau_k$ do not induce much concentration fluctuations.
If, however, one wished to decline this algorithm to study the stochastic fluctuation of the chemical reactions, it would be enough to use the relation $\tau_k= \frac{1}{\Omega\;v_k\left( S \right) }\log \left( \frac{1}{ U_k} \right) $ with   independent uniform random deviates $U_k$ in $[0,1]$ for the reaction time computations.
In the latter case, a realistic estimate for the $\Omega$ parameter value must be used. 

\section{Methods}

\subsection{Code availability and computer simulation}
\label{sec:numerical}
Both methods, SSA and DSA, were implemented with the same highest level of optimization, using low-level bit-manipulation tools to implement addressing operators in a Fortran code (compiled with gfortran and optimization flag ``-O3'').
Simulations were run on a standard laptop with an Intel(R) Core(TM) i5-6300U CPU at 2.40GHz. No paralellization were used. The fortran code is available in github\\ \url{https://github.com/Qthommen/Stochastic-method-for-isotope-labeling-systems.git}

\subsection{Goodness of Fit}
The chi-square per degree of freedom
$\chi^2_\nu=\frac{1}{n-p}\sum_{i=1}^n\frac{\left(y_i-y_i^*\right)^2}{\sigma_i^2}$
is used a goodness of fit criterion. $n$ is the number of targets; $p$, the number of parameters; $y_i$ et $y_i^*$  the computations and tagets; $\sigma_i^2$ the variance. The fit is accepted $\chi^2_\nu<1$.

\subsection{Metabolic Network}
\label{sec:chem_net}

{  Table~1} lists the chemical reactions and carbon rearrangements of the upper part of the increased glycolysis of the pentose phosphate pathway (Figure \ref{fig:video}). To illustrate the dynamics of the propagation of the labeled carbons and for the sake of simplicity, the reaction rate used corresponds to the mass action law with a kinetic parameter of unit value.

\begin{sidewaystable}
    
  \centering
\begin{tabular}{c|c|c|c}
reac nb & chemical reaction & reaction speed & parameter values \\ \hline\hline
    1
       & \schemestart GLU (abcde) \arrow{->} G6P (abcde) \schemestop
                           & $v_{G6P}=k_1$
                                   & $k_1=1 \rightarrow 2$ $\mu$M/s \\ 
    2
       & \schemestart G6P (abcdef)  \arrow{<=>} F6P (abcdef) \schemestop
                           & $v_2=k_2^+\,[G6P]-k_2^-\,[F6P]$
                                   & $k_2^+=k_2^-=1$/s\\
    3
       &\schemestart  F6P (abcdef) \arrow{<=>} FBP (abcdef) \schemestop
                           & $v_3=k_3^+[F6P]-k_3^-[FBP]$
                                   & $k_3^+=k_3^-=1$~/s \\
    4
       & \schemestart FBP (acbdef) \arrow{<=>} DHAP (cba) + GAP (def) \schemestop
                           & $v_4=k_4^+[FBP] - k_4^-[DHAP][GAP]$
                                   & $k_4^+=1$/s; $k_4^-=1$/$\mu$M/s \\
    5
       & \schemestart DHAP (abc) \arrow{<=>} GAP (abc) \schemestop
                           & $v_5=k_5^+[DHAP]-k_5^-[GAP]$
                                   & $k_5^+=k_5^-=1$~/s \\
    6
       & \schemestart G6P (abcdef) \arrow{->} 6PG (abcdef) \schemestop
                           & $v_6=k_6[G6P]$
                                   & $k_6=1$/s \\
    7
       & \schemestart 6PG (abcdef) \arrow{->} CO2 (a) + Ru5 (bcdef)  \schemestop
                           & $v_7=k_7[6PG]$
                                   & $k_7=1$/s \\
    8
       & \schemestart Ru5 (abcde) \arrow{<=>} R5P (abcde) \schemestop
                           & $v_8=k_8^+[Ru5]-k_8^-[R5P]$
                                   & $k_8^+=k_8^-=1$~/s \\
    9
       & \schemestart Ru5 (abcde) \arrow{<=>} X5P (abcde) \schemestop
                           & $v_9= k_9^+[Ru5]-k_9^-[X5P]$
                                   & $k_9^+=k_9^-=1$~/s \\
    10
       & \schemestart X5P (abcde) + R5P (ABCDE) \arrow{<=>} S7P (abABCDE) + GAP (cde) \schemestop
                           & $v_{10}=k_{10}^+[X5P][R5P] - k_{10}^-[S7P][GAP]$
                             & $k_{10}^+=k_{10}^-=1$~/$\mu$M/s \\
    11
       & \schemestart S7P (abcdefg) + GAP (ABC) \arrow{<=>} F6P (abcABC) + E4P (defg) \schemestop
                           & $v_{11}=k_{11}^+[S7P][GAP] - k_{11}^-[F6P][E4P]$
                             & $k_{11}^+=k_{11}^-=1$~/$\mu$M/s \\
    12
       & \schemestart X5P (abcde) + E4P (ABCD) \arrow{<=>} F6P (abABCD) + GAP (cde) \schemestop
                           & $v_{12}=k_{12}^+[X5P][E4P] - k_{12}^-[F6P][GAP]$
                                   & $k_{12}^+=k_{12}^-=1$~/$\mu$M/s \\
    13
       & \schemestart GAP (abc) \arrow{->} \schemestop
                           & $v_{13}=k_{13}[GAP]$
                                   & $k_{13}=1$~/s \\
    14
       & \schemestart R5P (abcde) \arrow{->} \schemestop
                           & $v_{14}=k_{14}[R5P]$
                                   & $k_{14}=1$~/s \\
    15
       & \schemestart CO2 (a) \arrow{->} \schemestop
                           & $v_{15}=k_{15}[CO2]$
                                   & $k_{15}=1$~/s \\
  \end{tabular}
  \caption{List of chemical reactions and reaction speed used to simulate the 13C propagation through the upper glycolytic pathways supplemented by the pentose
phosphate pathway displayed in Figure~1}
  \label{tab:sup_ppp_net}
\end{sidewaystable}

\clearpage
\section{Discussion}

In this study, we propose a stochastic algorithm to emulate the propagation of labeled atoms in a nonstationary metabolic system.
This algorithm derives from the chemical master equation which is the most comprehensive framework for describing chemical reaction network dynamics.
The efficiency of the algorithm has been applied to 13C-DMFA of the illustrative case of the pentose phosphate pathways for which 13C-labeling and concentration time series data has been synthesized.
One of the main computational advantages of the proposed method lies in the very weak dependence of the computation time on the length of the marking chain and thus the number of isotopomer.
Deterministic methods exhibit by construction a number of variables and a computation time that both rapidly increase with the combinatoriality associated to the power-law dependence with the chain length.
SSA is therefore well adapted to the study of parallel labeling, combining for instance carbon and hydrogen labeling~\citep{lewis2014tracing,antoniewicz2015parallel,Jacobson2019,Dong2019}.
Moreover, a simulation that mimics the stochastic and discrete nature of metabolic reaction processes provides a more accurate and comprehensive picture relating the propagation of labeling with the dynamics of isotopomer and metabolite concentrations.
Finally, this rigorous and straightforward method requires no tinkering or approximations depending on the resolution of the experimental measurements or the nature of the metabolic process, as it calculates all isotopomers at no additional cost and natively handles both stationary and non-stationary conditions.
In other words, the SSA method can be used interchangeably or simultaneously for 13C-MFA~\citep{Hurbain2022}, 13C-NMFA or 13C-DMFA. 
Because problem-dependent reduction or solving techniques are not used, the implementation does not require any particular software and can simply be done in any programming language, as it is the case for chemical kinetics modeling (see Code availability).

Estimation of metabolic flux dynamics from 13C labeling and metabolomics data can be done either by inverse kinetic model modeling~\citep{Wahl2008,baxter2007} or by considering flux function~\citep{antoniewicz2007DMFA,leighty2011DMFA,schumacher2015,quek2020}.
The preference of latter methods have been motivated by the lack of information about intracellular enzyme kinetics, but also the computational cost of deterministic simulation of kinetic models comprising isotopomer variables.
Thanks to the computational efficiency of SSA for simulating isotopomer dynamics, inverse kinetic modeling integrating 13C labeling data become an achievable goal. 
However, SSA can still be used with flux function as well, for instance with constant function in case of stationary metabolic condition (\textit{e.g.}, 13C-MFA and 13C-NMFA)~\citep{Hurbain2022}.
To summarize, a SSA-based 13C-DMFA method would require (1) defining a stoichiometry model, (2) defining kinetic laws or flux functions, (3) using an optimization method to estimate the parameters of reaction laws or flux functions, (4) using a Monte-Carlo method to evaluate the distribution of such parameters, (5) adjust iteratively the model (stoichiometry or kinetic structure) to optimize tradeoff between a good fit and a narrow parameter distributions. The last step corresponds to the well-known problem of model selection~\citep{mangan2017model}.
It is, however, to keep in mind that overparametrization is not a issue as long as one focuses on the estimation of flux trajectories. 
If, on the other hand, the 13C-DMFA is used for dynamic control purposes ~\citep{hartline2021dynamic}, the parameterization of flux functions will be of great importance, and it will be necessary to model the chemical kinetics as precisely as possible. 

The only delicate issue associated to this method is associated to the appropriate choice of the sample size $\Omega$.
$\Omega$ must be large enough to ensure that the level of fluctuations in isotopomer concentration are below the experimental uncertainties. 
At the same time, computational time scales linearly with $\Omega$ motivating to keep its value as low as possible.
The parameter $\Omega$ thus needs to be adjusted to a typical value (typically $100-1000$) to optimize the tradeoff between simulation uncertainties (below experimental uncertainties) and computational efficiency.
For such system size, the residual fluctuations of isotopomer concentration leads to a narrow distribution of error score for a same parameter set, which is not an issue when using Monte Carlo sampling algorithm used for metabolic flux analysis ~\citep{theorell2017,valderrama2019,heinonen2019,theorell2020}.
For a given value $\Omega$, a temporal averaging procedure may be added to narrow the distribution of mass isotopomer concentrations for given $\Omega$, allowing to use lower $\Omega$ values.
Another limitation relates to the high number of reactions which depends on the absolute value of directional fluxes, not of net fluxes.
This limitation can be largely compensated by the property that the number of operations (\textit{e.g.}, computational time) does not depend on isotopomer number per metabolites.


\section*{Acknowledgements}
The authors thank Darka Labavic for the fruitful discussions.

\section*{Funding}
This work has been supported by the LABEX CEMPI (ANR-11-LABX-0007) and by the Ministry of Higher Education and Research, Hauts de France council and European Regional Development Fund (ERDF) through the Contrat de Projets Etat-Region (CPER Photonics for Society P4S and CPER Cancer 2015-2020).

%
%


\begin{thebibliography}{41}%
\makeatletter
\providecommand \@ifxundefined [1]{%
 \@ifx{#1\undefined}
}%
\providecommand \@ifnum [1]{%
 \ifnum #1\expandafter \@firstoftwo
 \else \expandafter \@secondoftwo
 \fi
}%
\providecommand \@ifx [1]{%
 \ifx #1\expandafter \@firstoftwo
 \else \expandafter \@secondoftwo
 \fi
}%
\providecommand \natexlab [1]{#1}%
\providecommand \enquote  [1]{``#1''}%
\providecommand \bibnamefont  [1]{#1}%
\providecommand \bibfnamefont [1]{#1}%
\providecommand \citenamefont [1]{#1}%
\providecommand \href@noop [0]{\@secondoftwo}%
\providecommand \href [0]{\begingroup \@sanitize@url \@href}%
\providecommand \@href[1]{\@@startlink{#1}\@@href}%
\providecommand \@@href[1]{\endgroup#1\@@endlink}%
\providecommand \@sanitize@url [0]{\catcode `\\12\catcode `\$12\catcode
  `\&12\catcode `\#12\catcode `\^12\catcode `\_12\catcode `\%12\relax}%
\providecommand \@@startlink[1]{}%
\providecommand \@@endlink[0]{}%
\providecommand \url  [0]{\begingroup\@sanitize@url \@url }%
\providecommand \@url [1]{\endgroup\@href {#1}{\urlprefix }}%
\providecommand \urlprefix  [0]{URL }%
\providecommand \Eprint [0]{\href }%
\providecommand \doibase [0]{http://dx.doi.org/}%
\providecommand \selectlanguage [0]{\@gobble}%
\providecommand \bibinfo  [0]{\@secondoftwo}%
\providecommand \bibfield  [0]{\@secondoftwo}%
\providecommand \translation [1]{[#1]}%
\providecommand \BibitemOpen [0]{}%
\providecommand \bibitemStop [0]{}%
\providecommand \bibitemNoStop [0]{.\EOS\space}%
\providecommand \EOS [0]{\spacefactor3000\relax}%
\providecommand \BibitemShut  [1]{\csname bibitem#1\endcsname}%
\let\auto@bib@innerbib\@empty
\bibitem [{\citenamefont {Allen}\ and\ \citenamefont
  {Young}(2020)}]{allen2020}%
  \BibitemOpen
  \bibfield  {author} {\bibinfo {author} {\bibnamefont {Allen}, \bibfnamefont
  {D.~K.}}\ and\ \bibinfo {author} {\bibnamefont {Young}, \bibfnamefont
  {J.~D.}},\ }\bibfield  {title} {\enquote {\bibinfo {title} {Tracing metabolic
  flux through time and space with isotope labeling experiments},}\ }\href@noop
  {} {\bibfield  {journal} {\bibinfo  {journal} {Current opinion in
  biotechnology}\ }\textbf {\bibinfo {volume} {64}},\ \bibinfo {pages}
  {92--100} (\bibinfo {year} {2020})}\BibitemShut {NoStop}%
\bibitem [{\citenamefont {Antoniewicz}(2015{\natexlab{a}})}]{antoniewicz2015}%
  \BibitemOpen
  \bibfield  {author} {\bibinfo {author} {\bibnamefont {Antoniewicz},
  \bibfnamefont {M.~R.}},\ }\bibfield  {title} {\enquote {\bibinfo {title}
  {Methods and advances in metabolic flux analysis: a mini-review},}\
  }\href@noop {} {\bibfield  {journal} {\bibinfo  {journal} {Journal of
  industrial microbiology and biotechnology}\ }\textbf {\bibinfo {volume}
  {42}},\ \bibinfo {pages} {317--325} (\bibinfo {year}
  {2015}{\natexlab{a}})}\BibitemShut {NoStop}%
\bibitem [{\citenamefont
  {Antoniewicz}(2015{\natexlab{b}})}]{antoniewicz2015parallel}%
  \BibitemOpen
  \bibfield  {author} {\bibinfo {author} {\bibnamefont {Antoniewicz},
  \bibfnamefont {M.~R.}},\ }\bibfield  {title} {\enquote {\bibinfo {title}
  {Parallel labeling experiments for pathway elucidation and 13c metabolic flux
  analysis},}\ }\href@noop {} {\bibfield  {journal} {\bibinfo  {journal}
  {Current opinion in biotechnology}\ }\textbf {\bibinfo {volume} {36}},\
  \bibinfo {pages} {91--97} (\bibinfo {year} {2015}{\natexlab{b}})}\BibitemShut
  {NoStop}%
\bibitem [{\citenamefont {Antoniewicz}(2021)}]{antoniewicz2021guide}%
  \BibitemOpen
  \bibfield  {author} {\bibinfo {author} {\bibnamefont {Antoniewicz},
  \bibfnamefont {M.~R.}},\ }\bibfield  {title} {\enquote {\bibinfo {title} {A
  guide to metabolic flux analysis in metabolic engineering: methods, tools and
  applications},}\ }\href@noop {} {\bibfield  {journal} {\bibinfo  {journal}
  {Metabolic engineering}\ }\textbf {\bibinfo {volume} {63}},\ \bibinfo {pages}
  {2--12} (\bibinfo {year} {2021})}\BibitemShut {NoStop}%
\bibitem [{\citenamefont {Antoniewicz}, \citenamefont {Kelleher},\ and\
  \citenamefont {Stephanopoulos}(2007)}]{antoniewicz2007emu}%
  \BibitemOpen
  \bibfield  {author} {\bibinfo {author} {\bibnamefont {Antoniewicz},
  \bibfnamefont {M.~R.}}, \bibinfo {author} {\bibnamefont {Kelleher},
  \bibfnamefont {J.~K.}}, \ and\ \bibinfo {author} {\bibnamefont
  {Stephanopoulos}, \bibfnamefont {G.}},\ }\bibfield  {title} {\enquote
  {\bibinfo {title} {Elementary metabolite units ({EMU}): a novel framework for
  modeling isotopic distributions},}\ }\href@noop {} {\bibfield  {journal}
  {\bibinfo  {journal} {Metabolic engineering}\ }\textbf {\bibinfo {volume}
  {9}},\ \bibinfo {pages} {68--86} (\bibinfo {year} {2007})}\BibitemShut
  {NoStop}%
\bibitem [{\citenamefont {Antoniewicz}\ \emph {et~al.}(2007)\citenamefont
  {Antoniewicz}, \citenamefont {Kraynie}, \citenamefont {Laffend},
  \citenamefont {Gonz{\'a}lez-Lergier}, \citenamefont {Kelleher},\ and\
  \citenamefont {Stephanopoulos}}]{antoniewicz2007DMFA}%
  \BibitemOpen
  \bibfield  {author} {\bibinfo {author} {\bibnamefont {Antoniewicz},
  \bibfnamefont {M.~R.}}, \bibinfo {author} {\bibnamefont {Kraynie},
  \bibfnamefont {D.~F.}}, \bibinfo {author} {\bibnamefont {Laffend},
  \bibfnamefont {L.~A.}}, \bibinfo {author} {\bibnamefont
  {Gonz{\'a}lez-Lergier}, \bibfnamefont {J.}}, \bibinfo {author} {\bibnamefont
  {Kelleher}, \bibfnamefont {J.~K.}}, \ and\ \bibinfo {author} {\bibnamefont
  {Stephanopoulos}, \bibfnamefont {G.}},\ }\bibfield  {title} {\enquote
  {\bibinfo {title} {Metabolic flux analysis in a nonstationary system:
  fed-batch fermentation of a high yielding strain of {E}. coli producing 1,
  3-propanediol},}\ }\href@noop {} {\bibfield  {journal} {\bibinfo  {journal}
  {Metabolic engineering}\ }\textbf {\bibinfo {volume} {9}},\ \bibinfo {pages}
  {277--292} (\bibinfo {year} {2007})}\BibitemShut {NoStop}%
\bibitem [{\citenamefont {Baxter}\ \emph {et~al.}(2007)\citenamefont {Baxter},
  \citenamefont {Liu}, \citenamefont {Fernie},\ and\ \citenamefont
  {Sweetlove}}]{baxter2007}%
  \BibitemOpen
  \bibfield  {author} {\bibinfo {author} {\bibnamefont {Baxter}, \bibfnamefont
  {C.}}, \bibinfo {author} {\bibnamefont {Liu}, \bibfnamefont {J.}}, \bibinfo
  {author} {\bibnamefont {Fernie}, \bibfnamefont {A.}}, \ and\ \bibinfo
  {author} {\bibnamefont {Sweetlove}, \bibfnamefont {L.}},\ }\bibfield  {title}
  {\enquote {\bibinfo {title} {Determination of metabolic fluxes in a
  non-steady-state system},}\ }\href@noop {} {\bibfield  {journal} {\bibinfo
  {journal} {Phytochemistry}\ }\textbf {\bibinfo {volume} {68}},\ \bibinfo
  {pages} {2313--2319} (\bibinfo {year} {2007})}\BibitemShut {NoStop}%
\bibitem [{\citenamefont {Bouzier-Sore}\ and\ \citenamefont
  {Bola{\~n}os}(2015)}]{bouzier2015}%
  \BibitemOpen
  \bibfield  {author} {\bibinfo {author} {\bibnamefont {Bouzier-Sore},
  \bibfnamefont {A.-K.}}\ and\ \bibinfo {author} {\bibnamefont {Bola{\~n}os},
  \bibfnamefont {J.~P.}},\ }\bibfield  {title} {\enquote {\bibinfo {title}
  {Uncertainties in pentose-phosphate pathway flux assessment underestimate its
  contribution to neuronal glucose consumption: relevance for neurodegeneration
  and aging},}\ }\href@noop {} {\bibfield  {journal} {\bibinfo  {journal}
  {Frontiers in Aging Neuroscience}\ }\textbf {\bibinfo {volume} {7}},\
  \bibinfo {pages} {89} (\bibinfo {year} {2015})}\BibitemShut {NoStop}%
\bibitem [{\citenamefont {Creek}\ \emph {et~al.}(2015)\citenamefont {Creek},
  \citenamefont {Mazet}, \citenamefont {Achcar}, \citenamefont {Anderson},
  \citenamefont {Kim}, \citenamefont {Kamour}, \citenamefont {Morand},
  \citenamefont {Millerioux}, \citenamefont {Biran}, \citenamefont {Kerkhoven}
  \emph {et~al.}}]{creek2015}%
  \BibitemOpen
  \bibfield  {author} {\bibinfo {author} {\bibnamefont {Creek}, \bibfnamefont
  {D.~J.}}, \bibinfo {author} {\bibnamefont {Mazet}, \bibfnamefont {M.}},
  \bibinfo {author} {\bibnamefont {Achcar}, \bibfnamefont {F.}}, \bibinfo
  {author} {\bibnamefont {Anderson}, \bibfnamefont {J.}}, \bibinfo {author}
  {\bibnamefont {Kim}, \bibfnamefont {D.-H.}}, \bibinfo {author} {\bibnamefont
  {Kamour}, \bibfnamefont {R.}}, \bibinfo {author} {\bibnamefont {Morand},
  \bibfnamefont {P.}}, \bibinfo {author} {\bibnamefont {Millerioux},
  \bibfnamefont {Y.}}, \bibinfo {author} {\bibnamefont {Biran}, \bibfnamefont
  {M.}}, \bibinfo {author} {\bibnamefont {Kerkhoven}, \bibfnamefont {E.~J.}},
  \emph {et~al.},\ }\bibfield  {title} {\enquote {\bibinfo {title} {Probing the
  metabolic network in bloodstream-form trypanosoma brucei using untargeted
  metabolomics with stable isotope labelled glucose},}\ }\href@noop {}
  {\bibfield  {journal} {\bibinfo  {journal} {PLoS pathogens}\ }\textbf
  {\bibinfo {volume} {11}},\ \bibinfo {pages} {e1004689} (\bibinfo {year}
  {2015})}\BibitemShut {NoStop}%
\bibitem [{\citenamefont {Crown}\ and\ \citenamefont
  {Antoniewicz}(2013)}]{crown2013parallel}%
  \BibitemOpen
  \bibfield  {author} {\bibinfo {author} {\bibnamefont {Crown}, \bibfnamefont
  {S.~B.}}\ and\ \bibinfo {author} {\bibnamefont {Antoniewicz}, \bibfnamefont
  {M.~R.}},\ }\bibfield  {title} {\enquote {\bibinfo {title} {Parallel labeling
  experiments and metabolic flux analysis: Past, present and future
  methodologies},}\ }\href@noop {} {\bibfield  {journal} {\bibinfo  {journal}
  {Metabolic engineering}\ }\textbf {\bibinfo {volume} {16}},\ \bibinfo {pages}
  {21--32} (\bibinfo {year} {2013})}\BibitemShut {NoStop}%
\bibitem [{\citenamefont {Diaz-Moralli}\ \emph {et~al.}(2016)\citenamefont
  {Diaz-Moralli}, \citenamefont {Aguilar}, \citenamefont {Marin}, \citenamefont
  {Coy}, \citenamefont {Dewerchin}, \citenamefont {Antoniewicz}, \citenamefont
  {Meca-Cort{\'e}s}, \citenamefont {Notebaert}, \citenamefont {Ghesqui{\`e}re},
  \citenamefont {Eelen} \emph {et~al.}}]{diaz2016}%
  \BibitemOpen
  \bibfield  {author} {\bibinfo {author} {\bibnamefont {Diaz-Moralli},
  \bibfnamefont {S.}}, \bibinfo {author} {\bibnamefont {Aguilar}, \bibfnamefont
  {E.}}, \bibinfo {author} {\bibnamefont {Marin}, \bibfnamefont {S.}}, \bibinfo
  {author} {\bibnamefont {Coy}, \bibfnamefont {J.~F.}}, \bibinfo {author}
  {\bibnamefont {Dewerchin}, \bibfnamefont {M.}}, \bibinfo {author}
  {\bibnamefont {Antoniewicz}, \bibfnamefont {M.~R.}}, \bibinfo {author}
  {\bibnamefont {Meca-Cort{\'e}s}, \bibfnamefont {O.}}, \bibinfo {author}
  {\bibnamefont {Notebaert}, \bibfnamefont {L.}}, \bibinfo {author}
  {\bibnamefont {Ghesqui{\`e}re}, \bibfnamefont {B.}}, \bibinfo {author}
  {\bibnamefont {Eelen}, \bibfnamefont {G.}},  \emph {et~al.},\ }\bibfield
  {title} {\enquote {\bibinfo {title} {A key role for transketolase-like 1 in
  tumor metabolic reprogramming},}\ }\href@noop {} {\bibfield  {journal}
  {\bibinfo  {journal} {Oncotarget}\ }\textbf {\bibinfo {volume} {7}},\
  \bibinfo {pages} {51875} (\bibinfo {year} {2016})}\BibitemShut {NoStop}%
\bibitem [{\citenamefont {Dong}\ \emph {et~al.}(2019)\citenamefont {Dong},
  \citenamefont {Moon}, \citenamefont {Kelleher},\ and\ \citenamefont
  {Stephanopoulos}}]{Dong2019}%
  \BibitemOpen
  \bibfield  {author} {\bibinfo {author} {\bibnamefont {Dong}, \bibfnamefont
  {W.}}, \bibinfo {author} {\bibnamefont {Moon}, \bibfnamefont {S.~J.}},
  \bibinfo {author} {\bibnamefont {Kelleher}, \bibfnamefont {J.~K.}}, \ and\
  \bibinfo {author} {\bibnamefont {Stephanopoulos}, \bibfnamefont {G.}},\
  }\bibfield  {title} {\enquote {\bibinfo {title} {Dissecting mammalian cell
  metabolism through {13C}-and {2H}-isotope tracing: Interpretations at the
  molecular and systems levels},}\ }\href@noop {} {\bibfield  {journal}
  {\bibinfo  {journal} {Industrial \& Engineering Chemistry Research}\ }\textbf
  {\bibinfo {volume} {59}},\ \bibinfo {pages} {2593--2610} (\bibinfo {year}
  {2019})}\BibitemShut {NoStop}%
\bibitem [{\citenamefont {Gibson}\ and\ \citenamefont
  {Bruck}(2000)}]{gibson2000}%
  \BibitemOpen
  \bibfield  {author} {\bibinfo {author} {\bibnamefont {Gibson}, \bibfnamefont
  {M.~A.}}\ and\ \bibinfo {author} {\bibnamefont {Bruck}, \bibfnamefont {J.}},\
  }\bibfield  {title} {\enquote {\bibinfo {title} {Efficient exact stochastic
  simulation of chemical systems with many species and many channels},}\
  }\href@noop {} {\bibfield  {journal} {\bibinfo  {journal} {The journal of
  physical chemistry A}\ }\textbf {\bibinfo {volume} {104}},\ \bibinfo {pages}
  {1876--1889} (\bibinfo {year} {2000})}\BibitemShut {NoStop}%
\bibitem [{\citenamefont {Gillespie}(1977)}]{gillespie1977exact}%
  \BibitemOpen
  \bibfield  {author} {\bibinfo {author} {\bibnamefont {Gillespie},
  \bibfnamefont {D.~T.}},\ }\bibfield  {title} {\enquote {\bibinfo {title}
  {Exact stochastic simulation of coupled chemical reactions},}\ }\href@noop {}
  {\bibfield  {journal} {\bibinfo  {journal} {The journal of physical
  chemistry}\ }\textbf {\bibinfo {volume} {81}},\ \bibinfo {pages} {2340--2361}
  (\bibinfo {year} {1977})}\BibitemShut {NoStop}%
\bibitem [{\citenamefont {Gillespie}(1992)}]{Gillespie1992}%
  \BibitemOpen
  \bibfield  {author} {\bibinfo {author} {\bibnamefont {Gillespie},
  \bibfnamefont {D.~T.}},\ }\bibfield  {title} {\enquote {\bibinfo {title} {A
  rigorous derivation of the chemical master equation},}\ }\href@noop {}
  {\bibfield  {journal} {\bibinfo  {journal} {Physica A: Statistical Mechanics
  and its Applications}\ }\textbf {\bibinfo {volume} {188}},\ \bibinfo {pages}
  {404--425} (\bibinfo {year} {1992})}\BibitemShut {NoStop}%
\bibitem [{\citenamefont {Gillespie}(2000)}]{gillespie2000}%
  \BibitemOpen
  \bibfield  {author} {\bibinfo {author} {\bibnamefont {Gillespie},
  \bibfnamefont {D.~T.}},\ }\bibfield  {title} {\enquote {\bibinfo {title} {The
  chemical langevin equation},}\ }\href@noop {} {\bibfield  {journal} {\bibinfo
   {journal} {The Journal of Chemical Physics}\ }\textbf {\bibinfo {volume}
  {113}},\ \bibinfo {pages} {297--306} (\bibinfo {year} {2000})}\BibitemShut
  {NoStop}%
\bibitem [{\citenamefont {Gillespie}(2001)}]{gillespie2001approximate}%
  \BibitemOpen
  \bibfield  {author} {\bibinfo {author} {\bibnamefont {Gillespie},
  \bibfnamefont {D.~T.}},\ }\bibfield  {title} {\enquote {\bibinfo {title}
  {Approximate accelerated stochastic simulation of chemically reacting
  systems},}\ }\href@noop {} {\bibfield  {journal} {\bibinfo  {journal} {The
  Journal of chemical physics}\ }\textbf {\bibinfo {volume} {115}},\ \bibinfo
  {pages} {1716--1733} (\bibinfo {year} {2001})}\BibitemShut {NoStop}%
\bibitem [{\citenamefont {Hartline}\ \emph {et~al.}(2021)\citenamefont
  {Hartline}, \citenamefont {Schmitz}, \citenamefont {Han},\ and\ \citenamefont
  {Zhang}}]{hartline2021dynamic}%
  \BibitemOpen
  \bibfield  {author} {\bibinfo {author} {\bibnamefont {Hartline},
  \bibfnamefont {C.~J.}}, \bibinfo {author} {\bibnamefont {Schmitz},
  \bibfnamefont {A.~C.}}, \bibinfo {author} {\bibnamefont {Han}, \bibfnamefont
  {Y.}}, \ and\ \bibinfo {author} {\bibnamefont {Zhang}, \bibfnamefont {F.}},\
  }\bibfield  {title} {\enquote {\bibinfo {title} {Dynamic control in metabolic
  engineering: Theories, tools, and applications},}\ }\href@noop {} {\bibfield
  {journal} {\bibinfo  {journal} {Metabolic engineering}\ }\textbf {\bibinfo
  {volume} {63}},\ \bibinfo {pages} {126--140} (\bibinfo {year}
  {2021})}\BibitemShut {NoStop}%
\bibitem [{\citenamefont {Heinonen}\ \emph {et~al.}(2019)\citenamefont
  {Heinonen}, \citenamefont {Osmala}, \citenamefont {Mannerstr{\"o}m},
  \citenamefont {Wallenius}, \citenamefont {Kaski}, \citenamefont {Rousu},\
  and\ \citenamefont {L{\"a}hdesm{\"a}ki}}]{heinonen2019}%
  \BibitemOpen
  \bibfield  {author} {\bibinfo {author} {\bibnamefont {Heinonen},
  \bibfnamefont {M.}}, \bibinfo {author} {\bibnamefont {Osmala}, \bibfnamefont
  {M.}}, \bibinfo {author} {\bibnamefont {Mannerstr{\"o}m}, \bibfnamefont
  {H.}}, \bibinfo {author} {\bibnamefont {Wallenius}, \bibfnamefont {J.}},
  \bibinfo {author} {\bibnamefont {Kaski}, \bibfnamefont {S.}}, \bibinfo
  {author} {\bibnamefont {Rousu}, \bibfnamefont {J.}}, \ and\ \bibinfo {author}
  {\bibnamefont {L{\"a}hdesm{\"a}ki}, \bibfnamefont {H.}},\ }\bibfield  {title}
  {\enquote {\bibinfo {title} {Bayesian metabolic flux analysis reveals
  intracellular flux couplings},}\ }\href@noop {} {\bibfield  {journal}
  {\bibinfo  {journal} {Bioinformatics}\ }\textbf {\bibinfo {volume} {35}},\
  \bibinfo {pages} {i548--i557} (\bibinfo {year} {2019})}\BibitemShut {NoStop}%
\bibitem [{\citenamefont {Hurbain}\ \emph {et~al.}(2022)\citenamefont
  {Hurbain}, \citenamefont {Thommen}, \citenamefont {Anquez},\ and\
  \citenamefont {Pfeuty}}]{Hurbain2022}%
  \BibitemOpen
  \bibfield  {author} {\bibinfo {author} {\bibnamefont {Hurbain}, \bibfnamefont
  {J.}}, \bibinfo {author} {\bibnamefont {Thommen}, \bibfnamefont {Q.}},
  \bibinfo {author} {\bibnamefont {Anquez}, \bibfnamefont {F.}}, \ and\
  \bibinfo {author} {\bibnamefont {Pfeuty}, \bibfnamefont {B.}},\ }\bibfield
  {title} {\enquote {\bibinfo {title} {Quantitative modeling of pentose
  phosphate pathway response to oxidative stress reveals a cooperative
  regulatory strategy},}\ }\href {\doibase 10.1101/2022.02.04.478659}
  {\bibfield  {journal} {\bibinfo  {journal} {bioRxiv}\ } (\bibinfo {year}
  {2022}),\ 10.1101/2022.02.04.478659}\BibitemShut {NoStop}%
\bibitem [{\citenamefont {Jacobson}\ \emph {et~al.}(2019)\citenamefont
  {Jacobson}, \citenamefont {Adamczyk}, \citenamefont {Stevenson},
  \citenamefont {Regner}, \citenamefont {Ralph}, \citenamefont {Reed},\ and\
  \citenamefont {Amador-Noguez}}]{Jacobson2019}%
  \BibitemOpen
  \bibfield  {author} {\bibinfo {author} {\bibnamefont {Jacobson},
  \bibfnamefont {T.~B.}}, \bibinfo {author} {\bibnamefont {Adamczyk},
  \bibfnamefont {P.~A.}}, \bibinfo {author} {\bibnamefont {Stevenson},
  \bibfnamefont {D.~M.}}, \bibinfo {author} {\bibnamefont {Regner},
  \bibfnamefont {M.}}, \bibinfo {author} {\bibnamefont {Ralph}, \bibfnamefont
  {J.}}, \bibinfo {author} {\bibnamefont {Reed}, \bibfnamefont {J.~L.}}, \ and\
  \bibinfo {author} {\bibnamefont {Amador-Noguez}, \bibfnamefont {D.}},\
  }\bibfield  {title} {\enquote {\bibinfo {title} {{2H} and {13C} metabolic
  flux analysis elucidates in vivo thermodynamics of the ed pathway in
  zymomonas mobilis},}\ }\href@noop {} {\bibfield  {journal} {\bibinfo
  {journal} {Metabolic engineering}\ }\textbf {\bibinfo {volume} {54}},\
  \bibinfo {pages} {301--316} (\bibinfo {year} {2019})}\BibitemShut {NoStop}%
\bibitem [{\citenamefont {Kuehne}\ \emph {et~al.}(2015)\citenamefont {Kuehne},
  \citenamefont {Emmert}, \citenamefont {Soehle}, \citenamefont {Winnefeld},
  \citenamefont {Fischer}, \citenamefont {Wenck}, \citenamefont {Gallinat},
  \citenamefont {Terstegen}, \citenamefont {Lucius}, \citenamefont {Hildebrand}
  \emph {et~al.}}]{kuehne2015}%
  \BibitemOpen
  \bibfield  {author} {\bibinfo {author} {\bibnamefont {Kuehne}, \bibfnamefont
  {A.}}, \bibinfo {author} {\bibnamefont {Emmert}, \bibfnamefont {H.}},
  \bibinfo {author} {\bibnamefont {Soehle}, \bibfnamefont {J.}}, \bibinfo
  {author} {\bibnamefont {Winnefeld}, \bibfnamefont {M.}}, \bibinfo {author}
  {\bibnamefont {Fischer}, \bibfnamefont {F.}}, \bibinfo {author} {\bibnamefont
  {Wenck}, \bibfnamefont {H.}}, \bibinfo {author} {\bibnamefont {Gallinat},
  \bibfnamefont {S.}}, \bibinfo {author} {\bibnamefont {Terstegen},
  \bibfnamefont {L.}}, \bibinfo {author} {\bibnamefont {Lucius}, \bibfnamefont
  {R.}}, \bibinfo {author} {\bibnamefont {Hildebrand}, \bibfnamefont {J.}},
  \emph {et~al.},\ }\bibfield  {title} {\enquote {\bibinfo {title} {Acute
  activation of oxidative pentose phosphate pathway as first-line response to
  oxidative stress in human skin cells},}\ }\href@noop {} {\bibfield  {journal}
  {\bibinfo  {journal} {Molecular cell}\ }\textbf {\bibinfo {volume} {59}},\
  \bibinfo {pages} {359--371} (\bibinfo {year} {2015})}\BibitemShut {NoStop}%
\bibitem [{\citenamefont {Lee}\ \emph {et~al.}(2019)\citenamefont {Lee},
  \citenamefont {Malloy}, \citenamefont {Corbin}, \citenamefont {Li},\ and\
  \citenamefont {Jin}}]{lee2019}%
  \BibitemOpen
  \bibfield  {author} {\bibinfo {author} {\bibnamefont {Lee}, \bibfnamefont
  {M.~H.}}, \bibinfo {author} {\bibnamefont {Malloy}, \bibfnamefont {C.~R.}},
  \bibinfo {author} {\bibnamefont {Corbin}, \bibfnamefont {I.~R.}}, \bibinfo
  {author} {\bibnamefont {Li}, \bibfnamefont {J.}}, \ and\ \bibinfo {author}
  {\bibnamefont {Jin}, \bibfnamefont {E.~S.}},\ }\bibfield  {title} {\enquote
  {\bibinfo {title} {Assessing the pentose phosphate pathway using [2, 3-13c2]
  glucose},}\ }\href@noop {} {\bibfield  {journal} {\bibinfo  {journal} {NMR in
  Biomedicine}\ }\textbf {\bibinfo {volume} {32}},\ \bibinfo {pages} {e4096}
  (\bibinfo {year} {2019})}\BibitemShut {NoStop}%
\bibitem [{\citenamefont {Leighty}\ and\ \citenamefont
  {Antoniewicz}(2011)}]{leighty2011DMFA}%
  \BibitemOpen
  \bibfield  {author} {\bibinfo {author} {\bibnamefont {Leighty}, \bibfnamefont
  {R.~W.}}\ and\ \bibinfo {author} {\bibnamefont {Antoniewicz}, \bibfnamefont
  {M.~R.}},\ }\bibfield  {title} {\enquote {\bibinfo {title} {Dynamic metabolic
  flux analysis ({DMFA}): a framework for determining fluxes at metabolic
  non-steady state},}\ }\href@noop {} {\bibfield  {journal} {\bibinfo
  {journal} {Metabolic engineering}\ }\textbf {\bibinfo {volume} {13}},\
  \bibinfo {pages} {745--755} (\bibinfo {year} {2011})}\BibitemShut {NoStop}%
\bibitem [{\citenamefont {Lewis}\ \emph {et~al.}(2014)\citenamefont {Lewis},
  \citenamefont {Parker}, \citenamefont {Fiske}, \citenamefont {McCloskey},
  \citenamefont {Gui}, \citenamefont {Green}, \citenamefont {Vokes},
  \citenamefont {Feist}, \citenamefont {Vander~Heiden},\ and\ \citenamefont
  {Metallo}}]{lewis2014tracing}%
  \BibitemOpen
  \bibfield  {author} {\bibinfo {author} {\bibnamefont {Lewis}, \bibfnamefont
  {C.~A.}}, \bibinfo {author} {\bibnamefont {Parker}, \bibfnamefont {S.~J.}},
  \bibinfo {author} {\bibnamefont {Fiske}, \bibfnamefont {B.~P.}}, \bibinfo
  {author} {\bibnamefont {McCloskey}, \bibfnamefont {D.}}, \bibinfo {author}
  {\bibnamefont {Gui}, \bibfnamefont {D.~Y.}}, \bibinfo {author} {\bibnamefont
  {Green}, \bibfnamefont {C.~R.}}, \bibinfo {author} {\bibnamefont {Vokes},
  \bibfnamefont {N.~I.}}, \bibinfo {author} {\bibnamefont {Feist},
  \bibfnamefont {A.~M.}}, \bibinfo {author} {\bibnamefont {Vander~Heiden},
  \bibfnamefont {M.~G.}}, \ and\ \bibinfo {author} {\bibnamefont {Metallo},
  \bibfnamefont {C.~M.}},\ }\bibfield  {title} {\enquote {\bibinfo {title}
  {Tracing compartmentalized nadph metabolism in the cytosol and mitochondria
  of mammalian cells},}\ }\href@noop {} {\bibfield  {journal} {\bibinfo
  {journal} {Molecular cell}\ }\textbf {\bibinfo {volume} {55}},\ \bibinfo
  {pages} {253--263} (\bibinfo {year} {2014})}\BibitemShut {NoStop}%
\bibitem [{\citenamefont {Mangan}\ \emph {et~al.}(2017)\citenamefont {Mangan},
  \citenamefont {Kutz}, \citenamefont {Brunton},\ and\ \citenamefont
  {Proctor}}]{mangan2017model}%
  \BibitemOpen
  \bibfield  {author} {\bibinfo {author} {\bibnamefont {Mangan}, \bibfnamefont
  {N.~M.}}, \bibinfo {author} {\bibnamefont {Kutz}, \bibfnamefont {J.~N.}},
  \bibinfo {author} {\bibnamefont {Brunton}, \bibfnamefont {S.~L.}}, \ and\
  \bibinfo {author} {\bibnamefont {Proctor}, \bibfnamefont {J.~L.}},\
  }\bibfield  {title} {\enquote {\bibinfo {title} {Model selection for
  dynamical systems via sparse regression and information criteria},}\
  }\href@noop {} {\bibfield  {journal} {\bibinfo  {journal} {Proceedings of the
  Royal Society A: Mathematical, Physical and Engineering Sciences}\ }\textbf
  {\bibinfo {volume} {473}},\ \bibinfo {pages} {20170009} (\bibinfo {year}
  {2017})}\BibitemShut {NoStop}%
\bibitem [{\citenamefont {Metallo}, \citenamefont {Walther},\ and\
  \citenamefont {Stephanopoulos}(2009)}]{metallo2009}%
  \BibitemOpen
  \bibfield  {author} {\bibinfo {author} {\bibnamefont {Metallo}, \bibfnamefont
  {C.~M.}}, \bibinfo {author} {\bibnamefont {Walther}, \bibfnamefont {J.~L.}},
  \ and\ \bibinfo {author} {\bibnamefont {Stephanopoulos}, \bibfnamefont
  {G.}},\ }\bibfield  {title} {\enquote {\bibinfo {title} {Evaluation of {13C}
  isotopic tracers for metabolic flux analysis in mammalian cells},}\
  }\href@noop {} {\bibfield  {journal} {\bibinfo  {journal} {Journal of
  biotechnology}\ }\textbf {\bibinfo {volume} {144}},\ \bibinfo {pages}
  {167--174} (\bibinfo {year} {2009})}\BibitemShut {NoStop}%
\bibitem [{\citenamefont {Niedenf{\"u}hr}, \citenamefont {Wiechert},\ and\
  \citenamefont {N{\"o}h}(2015)}]{niedenfuhr2015}%
  \BibitemOpen
  \bibfield  {author} {\bibinfo {author} {\bibnamefont {Niedenf{\"u}hr},
  \bibfnamefont {S.}}, \bibinfo {author} {\bibnamefont {Wiechert},
  \bibfnamefont {W.}}, \ and\ \bibinfo {author} {\bibnamefont {N{\"o}h},
  \bibfnamefont {K.}},\ }\bibfield  {title} {\enquote {\bibinfo {title} {How to
  measure metabolic fluxes: a taxonomic guide for 13c fluxomics},}\ }\href@noop
  {} {\bibfield  {journal} {\bibinfo  {journal} {Current opinion in
  biotechnology}\ }\textbf {\bibinfo {volume} {34}},\ \bibinfo {pages} {82--90}
  (\bibinfo {year} {2015})}\BibitemShut {NoStop}%
\bibitem [{\citenamefont {Ohno}\ \emph {et~al.}(2020)\citenamefont {Ohno},
  \citenamefont {Quek}, \citenamefont {Krycer}, \citenamefont {Yugi},
  \citenamefont {Hirayama}, \citenamefont {Ikeda}, \citenamefont {Shoji},
  \citenamefont {Suzuki}, \citenamefont {Soga}, \citenamefont {James} \emph
  {et~al.}}]{ohno2020}%
  \BibitemOpen
  \bibfield  {author} {\bibinfo {author} {\bibnamefont {Ohno}, \bibfnamefont
  {S.}}, \bibinfo {author} {\bibnamefont {Quek}, \bibfnamefont {L.-E.}},
  \bibinfo {author} {\bibnamefont {Krycer}, \bibfnamefont {J.~R.}}, \bibinfo
  {author} {\bibnamefont {Yugi}, \bibfnamefont {K.}}, \bibinfo {author}
  {\bibnamefont {Hirayama}, \bibfnamefont {A.}}, \bibinfo {author}
  {\bibnamefont {Ikeda}, \bibfnamefont {S.}}, \bibinfo {author} {\bibnamefont
  {Shoji}, \bibfnamefont {F.}}, \bibinfo {author} {\bibnamefont {Suzuki},
  \bibfnamefont {K.}}, \bibinfo {author} {\bibnamefont {Soga}, \bibfnamefont
  {T.}}, \bibinfo {author} {\bibnamefont {James}, \bibfnamefont {D.~E.}},
  \emph {et~al.},\ }\bibfield  {title} {\enquote {\bibinfo {title} {Kinetic
  trans-omic analysis reveals key regulatory mechanisms for insulin-regulated
  glucose metabolism in adipocytes},}\ }\href@noop {} {\bibfield  {journal}
  {\bibinfo  {journal} {Iscience}\ }\textbf {\bibinfo {volume} {23}},\ \bibinfo
  {pages} {101479} (\bibinfo {year} {2020})}\BibitemShut {NoStop}%
\bibitem [{\citenamefont {Quek}\ \emph {et~al.}(2020)\citenamefont {Quek},
  \citenamefont {Krycer}, \citenamefont {Ohno}, \citenamefont {Yugi},
  \citenamefont {Fazakerley}, \citenamefont {Scalzo}, \citenamefont
  {Elkington}, \citenamefont {Dai}, \citenamefont {Hirayama}, \citenamefont
  {Ikeda} \emph {et~al.}}]{quek2020}%
  \BibitemOpen
  \bibfield  {author} {\bibinfo {author} {\bibnamefont {Quek}, \bibfnamefont
  {L.-E.}}, \bibinfo {author} {\bibnamefont {Krycer}, \bibfnamefont {J.~R.}},
  \bibinfo {author} {\bibnamefont {Ohno}, \bibfnamefont {S.}}, \bibinfo
  {author} {\bibnamefont {Yugi}, \bibfnamefont {K.}}, \bibinfo {author}
  {\bibnamefont {Fazakerley}, \bibfnamefont {D.~J.}}, \bibinfo {author}
  {\bibnamefont {Scalzo}, \bibfnamefont {R.}}, \bibinfo {author} {\bibnamefont
  {Elkington}, \bibfnamefont {S.~D.}}, \bibinfo {author} {\bibnamefont {Dai},
  \bibfnamefont {Z.}}, \bibinfo {author} {\bibnamefont {Hirayama},
  \bibfnamefont {A.}}, \bibinfo {author} {\bibnamefont {Ikeda}, \bibfnamefont
  {S.}},  \emph {et~al.},\ }\bibfield  {title} {\enquote {\bibinfo {title}
  {Dynamic {13C} flux analysis captures the reorganization of adipocyte glucose
  metabolism in response to insulin},}\ }\href@noop {} {\bibfield  {journal}
  {\bibinfo  {journal} {Iscience}\ }\textbf {\bibinfo {volume} {23}},\ \bibinfo
  {pages} {100855} (\bibinfo {year} {2020})}\BibitemShut {NoStop}%
\bibitem [{\citenamefont {Schmidt}\ \emph {et~al.}(1997)\citenamefont
  {Schmidt}, \citenamefont {Carlsen}, \citenamefont {Nielsen},\ and\
  \citenamefont {Villadsen}}]{schmidt1997idv}%
  \BibitemOpen
  \bibfield  {author} {\bibinfo {author} {\bibnamefont {Schmidt}, \bibfnamefont
  {K.}}, \bibinfo {author} {\bibnamefont {Carlsen}, \bibfnamefont {M.}},
  \bibinfo {author} {\bibnamefont {Nielsen}, \bibfnamefont {J.}}, \ and\
  \bibinfo {author} {\bibnamefont {Villadsen}, \bibfnamefont {J.}},\ }\bibfield
   {title} {\enquote {\bibinfo {title} {Modeling isotopomer distributions in
  biochemical networks using isotopomer mapping matrices},}\ }\href@noop {}
  {\bibfield  {journal} {\bibinfo  {journal} {Biotechnology and
  bioengineering}\ }\textbf {\bibinfo {volume} {55}},\ \bibinfo {pages}
  {831--840} (\bibinfo {year} {1997})}\BibitemShut {NoStop}%
\bibitem [{\citenamefont {Schumacher}\ and\ \citenamefont
  {Wahl}(2015)}]{schumacher2015}%
  \BibitemOpen
  \bibfield  {author} {\bibinfo {author} {\bibnamefont {Schumacher},
  \bibfnamefont {R.}}\ and\ \bibinfo {author} {\bibnamefont {Wahl},
  \bibfnamefont {S.~A.}},\ }\bibfield  {title} {\enquote {\bibinfo {title}
  {Effective estimation of dynamic metabolic fluxes using 13c labeling and
  piecewise affine approximation: from theory to practical applicability},}\
  }\href@noop {} {\bibfield  {journal} {\bibinfo  {journal} {Metabolites}\
  }\textbf {\bibinfo {volume} {5}},\ \bibinfo {pages} {697--719} (\bibinfo
  {year} {2015})}\BibitemShut {NoStop}%
\bibitem [{\citenamefont {Selivanov}\ \emph {et~al.}(2004)\citenamefont
  {Selivanov}, \citenamefont {Puigjaner}, \citenamefont {Sillero},
  \citenamefont {Centelles}, \citenamefont {Ramos-Montoya}, \citenamefont
  {Lee},\ and\ \citenamefont {Cascante}}]{selivanov2004}%
  \BibitemOpen
  \bibfield  {author} {\bibinfo {author} {\bibnamefont {Selivanov},
  \bibfnamefont {V.~A.}}, \bibinfo {author} {\bibnamefont {Puigjaner},
  \bibfnamefont {J.}}, \bibinfo {author} {\bibnamefont {Sillero}, \bibfnamefont
  {A.}}, \bibinfo {author} {\bibnamefont {Centelles}, \bibfnamefont {J.~J.}},
  \bibinfo {author} {\bibnamefont {Ramos-Montoya}, \bibfnamefont {A.}},
  \bibinfo {author} {\bibnamefont {Lee}, \bibfnamefont {P.~W.-N.}}, \ and\
  \bibinfo {author} {\bibnamefont {Cascante}, \bibfnamefont {M.}},\ }\bibfield
  {title} {\enquote {\bibinfo {title} {An optimized algorithm for flux
  estimation from isotopomer distribution in glucose metabolites},}\
  }\href@noop {} {\bibfield  {journal} {\bibinfo  {journal} {Bioinformatics}\
  }\textbf {\bibinfo {volume} {20}},\ \bibinfo {pages} {3387--3397} (\bibinfo
  {year} {2004})}\BibitemShut {NoStop}%
\bibitem [{\citenamefont {Stephanopoulos}(1999)}]{stephanopoulos1999metabolic}%
  \BibitemOpen
  \bibfield  {author} {\bibinfo {author} {\bibnamefont {Stephanopoulos},
  \bibfnamefont {G.}},\ }\bibfield  {title} {\enquote {\bibinfo {title}
  {Metabolic fluxes and metabolic engineering},}\ }\href@noop {} {\bibfield
  {journal} {\bibinfo  {journal} {Metabolic engineering}\ }\textbf {\bibinfo
  {volume} {1}},\ \bibinfo {pages} {1--11} (\bibinfo {year}
  {1999})}\BibitemShut {NoStop}%
\bibitem [{\citenamefont {Theorell}\ \emph {et~al.}(2017)\citenamefont
  {Theorell}, \citenamefont {Leweke}, \citenamefont {Wiechert},\ and\
  \citenamefont {N{\"o}h}}]{theorell2017}%
  \BibitemOpen
  \bibfield  {author} {\bibinfo {author} {\bibnamefont {Theorell},
  \bibfnamefont {A.}}, \bibinfo {author} {\bibnamefont {Leweke}, \bibfnamefont
  {S.}}, \bibinfo {author} {\bibnamefont {Wiechert}, \bibfnamefont {W.}}, \
  and\ \bibinfo {author} {\bibnamefont {N{\"o}h}, \bibfnamefont {K.}},\
  }\bibfield  {title} {\enquote {\bibinfo {title} {To be certain about the
  uncertainty: Bayesian statistics for 13c metabolic flux analysis},}\
  }\href@noop {} {\bibfield  {journal} {\bibinfo  {journal} {Biotechnology and
  bioengineering}\ }\textbf {\bibinfo {volume} {114}},\ \bibinfo {pages}
  {2668--2684} (\bibinfo {year} {2017})}\BibitemShut {NoStop}%
\bibitem [{\citenamefont {Theorell}\ and\ \citenamefont
  {N{\"o}h}(2020)}]{theorell2020}%
  \BibitemOpen
  \bibfield  {author} {\bibinfo {author} {\bibnamefont {Theorell},
  \bibfnamefont {A.}}\ and\ \bibinfo {author} {\bibnamefont {N{\"o}h},
  \bibfnamefont {K.}},\ }\bibfield  {title} {\enquote {\bibinfo {title}
  {Reversible jump {MCMC} for multi-model inference in metabolic flux
  analysis},}\ }\href@noop {} {\bibfield  {journal} {\bibinfo  {journal}
  {Bioinformatics}\ }\textbf {\bibinfo {volume} {36}},\ \bibinfo {pages}
  {232--240} (\bibinfo {year} {2020})}\BibitemShut {NoStop}%
\bibitem [{\citenamefont {Thompson}\ and\ \citenamefont
  {Stewart}(2002)}]{Thompson2002}%
  \BibitemOpen
  \bibfield  {author} {\bibinfo {author} {\bibnamefont {Thompson},
  \bibfnamefont {J.~M.~T.}}\ and\ \bibinfo {author} {\bibnamefont {Stewart},
  \bibfnamefont {H.~B.}},\ }\href@noop {} {\emph {\bibinfo {title} {Nonlinear
  dynamics and chaos}}}\ (\bibinfo  {publisher} {John Wiley \& Sons},\ \bibinfo
  {year} {2002})\BibitemShut {NoStop}%
\bibitem [{\citenamefont {Valderrama-Baham{\'o}ndez}\ and\ \citenamefont
  {Fr{\"o}hlich}(2019)}]{valderrama2019}%
  \BibitemOpen
  \bibfield  {author} {\bibinfo {author} {\bibnamefont
  {Valderrama-Baham{\'o}ndez}, \bibfnamefont {G.~I.}}\ and\ \bibinfo {author}
  {\bibnamefont {Fr{\"o}hlich}, \bibfnamefont {H.}},\ }\bibfield  {title}
  {\enquote {\bibinfo {title} {Mcmc techniques for parameter estimation of ode
  based models in systems biology},}\ }\href@noop {} {\bibfield  {journal}
  {\bibinfo  {journal} {Frontiers in Applied Mathematics and Statistics}\
  }\textbf {\bibinfo {volume} {5}},\ \bibinfo {pages} {55} (\bibinfo {year}
  {2019})}\BibitemShut {NoStop}%
\bibitem [{\citenamefont {Van~Kampen}(1992)}]{van1992}%
  \BibitemOpen
  \bibfield  {author} {\bibinfo {author} {\bibnamefont {Van~Kampen},
  \bibfnamefont {N.~G.}},\ }\href@noop {} {\emph {\bibinfo {title} {Stochastic
  processes in physics and chemistry}}},\ Vol.~\bibinfo {volume} {1}\ (\bibinfo
   {publisher} {Elsevier},\ \bibinfo {year} {1992})\BibitemShut {NoStop}%
\bibitem [{\citenamefont {Wahl}, \citenamefont {N{\"o}h},\ and\ \citenamefont
  {Wiechert}(2008)}]{Wahl2008}%
  \BibitemOpen
  \bibfield  {author} {\bibinfo {author} {\bibnamefont {Wahl}, \bibfnamefont
  {S.~A.}}, \bibinfo {author} {\bibnamefont {N{\"o}h}, \bibfnamefont {K.}}, \
  and\ \bibinfo {author} {\bibnamefont {Wiechert}, \bibfnamefont {W.}},\
  }\bibfield  {title} {\enquote {\bibinfo {title} {13 c labeling experiments at
  metabolic nonstationary conditions: an exploratory study},}\ }\href@noop {}
  {\bibfield  {journal} {\bibinfo  {journal} {Bmc Bioinformatics}\ }\textbf
  {\bibinfo {volume} {9}},\ \bibinfo {pages} {1--18} (\bibinfo {year}
  {2008})}\BibitemShut {NoStop}%
\bibitem [{\citenamefont {Zupke}\ and\ \citenamefont
  {Stephanopoulos}(1994)}]{zupke1994}%
  \BibitemOpen
  \bibfield  {author} {\bibinfo {author} {\bibnamefont {Zupke}, \bibfnamefont
  {C.}}\ and\ \bibinfo {author} {\bibnamefont {Stephanopoulos}, \bibfnamefont
  {G.}},\ }\bibfield  {title} {\enquote {\bibinfo {title} {Modeling of isotope
  distributions and intracellular fluxes in metabolic networks using atom
  mapping matrixes},}\ }\href@noop {} {\bibfield  {journal} {\bibinfo
  {journal} {Biotechnology Progress}\ }\textbf {\bibinfo {volume} {10}},\
  \bibinfo {pages} {489--498} (\bibinfo {year} {1994})}\BibitemShut {NoStop}%
\end{thebibliography}
%

\end{document}